\documentclass[10pt ]{revtex4}
\usepackage{hyperref}
\usepackage{amsmath,amsfonts}
\usepackage[dvips]{graphics}
\usepackage[dvips]{graphicx}
\usepackage{subfigure}
\usepackage{epsfig}

\RequirePackage{color}

\begin{document}

\title{Time dependent dark energy and the thermodynamics of many-body systems}
 \author{Behnam Pourhassan${}^{a, b}$}
  \email{b.pourhassan@du.ac.ir}

 \author{Sudhaker Upadhyay${}^{a, c, d}$}
  \email{sudhakerupadhyay@gmail.com}

 \affiliation{${}^{a}$School of Physics, Damghan University, P. O. Box 3671641167, Damghan, Iran}
  \affiliation{${}^{b}$ Canadian Quantum Research Center 204-3002 32 Ave Vernon, BC V1T 2L7 Canada}
 \affiliation{${}^{c}$Department of Physics, K. L. S. College, Magadh University, Nawada-805110,  India}
 \affiliation{${}^{d}$Visiting Associate, Inter-University Centre for Astronomy and Astrophysics (IUCAA) Pune, Maharashtra-411007}

\begin{abstract}
In this paper, we study the  thermodynamics and statistics of the  galaxies clustering affected by the dynamical dark energy. We consider two important dark energy models based on time dependent equation of state to evaluate the gravitational partition function. In the first model, we consider barotropic dark energy with time dependent equation of state. However,  in the second model, we consider various kinds of Chaplygin gas equation of state which originally introduced by string theory. We calculate, analytically and numerically, the thermodynamic  quantities in canonical and grand canonical ensembles. We investigate validity of the second law of thermodynamics for the total system of clustering of galaxies and dynamical dark energy. We finally evaluate the galaxy-galaxy correlation function and compare our model with Peebles's power law and find that the model based on generalized Chaplygin gas may yields to more agreement with observations.
\end{abstract}

\begin{keywords}
{Dark energy, Thermodynamics and Statistics, Clustering of Galaxies.}
\end{keywords}
 \maketitle

\section{Introduction}
Type Ia Supernovae (SNeIa) \cite{Riess} and the Cosmic Microwave Background (CMB) \cite{Jarosik} confirmed accelerating expansion of Universe,   but the physical
origin of this acceleration is unknown.   Theoretical solution for the accelerating expansion is presence of dark energy \cite{Bamba}.  In this regard, there are some theoretical models of dark energy. The simplest one is based on the cosmological constant $\Lambda$ which called $\Lambda$-CDM model \cite{Bahcall}. Because of the fine-tuning \cite{Copeland}  and the cosmic coincidence problems \cite{Nobbenhuis}, we need alternative model to describe dynamical dark energy. In that case, there are some dynamical model of dark energy based on the scalar field like phantom \cite{Caldwell}, quintessence \cite{Wetterich}, K-essence \cite{Armendariz-Picon} and tachyonic models \cite{Sen}. An interesting model of dynamical dark energy is based on Chaplygin gas equation of state \cite{Kamenshchik, Bento}. An important model of this case is called generalized Chaplygin gas \cite{Bilic}, or modified Chaplygin gas \cite{Debnath,Saadat2}, which is unified by the extended Chaplygin gas \cite{Kahya,Pourhassan}. Importance of Chaplygin gas models is unification of the dark energy and dark matter, which is firstly inspired by string theory point of view \cite{Bar1,Bar2}. It is possible to obtain Chaplygin gas equation of state from D-brane action of string theory \cite{O}. Holographic model of dark energy is another interesting model inspired by holographic entropy \cite{Li,Elizalde, SadPour}. A class of models, called
barotropic fluids, in which the dark energy pressure
is given as an explicit function of the density, have also been studied \cite{lind}.
Barotropic fluid models have a number of features that
make them an interesting class of dark energy models.
They have  an explicit   equation of state relating the pressure and the energy density.
Also, Modified theories of gravity are alternative to dark energy to explain accelerating expansion of the Universe like $f(R)$ theory \cite{Capozziello1,Capozziello2,Carroll,Khur}.
  We examine the class of barotropic fluid  and Chaplygin gas models of dark energy, in which the pressure is an
explicit function of the density.\\
As mentioned above, Chaplygin gas is interesting model of dark energy in particle physics, astrophysics and cosmology. In that case, modified Chaplygin gas \cite{Uraz} considered to study inflationary scenario \cite{sharif}, where the effects of bulk viscous pressure \cite{Saadat3} on warm inflationary modified Chaplygin gas model using Friedmann-Robertson-Walker background has been studied. It should be mention that bulk viscosity \cite{Pourhassan2} is important parameter in the energy momentum of a perfect baryotropic fluid \cite{Lidsey}.  Constraints on modified Chaplygin gas also investigated by \cite{Thakur}. Hence, all cosmological model based on Chaplygin gas including bulk viscosity unified through extended Chaplygin gas model \cite{Kahya2}, which can extracted from string theory, and these are our motivation to choose such dark energy model in part of this paper. For simplicity, we consider a toy model including  generalized Chaplygin gas which can be extended to better models of Chaplygin gas like modified or extended Chaplygin gas \cite{Adv}.\\
Recently, Ref. \cite{main} have been studied the effects of cosmological constant as dark energy on the thermodynamics of clustering of galaxies. Straightforward way to have a dynamical model of dark energy is consideration of varying $\Lambda$ \cite{Kahya,Khur,Jamil2}. Hence, \cite{PourMNRAS} have been studied thermodynamics of the clustering of galaxies under the effect of dynamical dark energy. Indeed varying $\Lambda$ as time dependent with power law considered as dark energy and correlation function of galaxies have been studied by Ref. \cite{PourMNRAS}. Now, we will consider more real model based on two important models of dark energy, and study the effect of time-dependent cosmological constant with combination of power law and exponential on the clustering of
galaxies.\\
Because of inter-galactic distances, it is possible to represent a system of galaxies as a point particle system. Hence one can use standard methods of statistical mechanics known as the method of cluster expansions to study clustering of galaxies \cite{ahm02}. In that case,  the effects of super-light brane world perturbative modes on structure formation has been studied  \cite{1}. Also, the galactic clustering of expanding Universe by assuming the gravitational interaction through the modified Newton's potential by means of the $f(R)$ gravity is investigated \cite{2}. It has been found that by  increasing   the field content of general relativity changes the large distance behavior of the theory \cite{3}. An interesting model of galaxies clustering discussed in the Ref. \cite{4}. Moreover, the thermodynamics of galactic clustering under the higher-order corrected Newtonian dynamics investigated by the Ref. \cite{5}. Indeed, the observed peculiar velocity distribution function, which is used to study the clustering parameters, has been considered for a sample of galaxies of order $50 Mpc$ for the local group \cite{1a}. Statistical mechanics prepare useful method to study clustering of galaxies. For example, it is possible to study spatial distribution function of galaxies at high redshift \cite{4a}, or to calculate the probability to have special shape of a galaxy clustering \cite{5a}. It is also possible to extend point like masses to the extended mass with finite size \cite{7a}. In this paper, we use the standard methods of statistical mechanics to study the clustering of galaxies in presence of barotropic fluid  and Chaplygin gas   dark energy models and compare the result. The main difference of our work with previous one like \cite{1,2,3,4,5} is the general forl of dark energy which will be considered in this paper.\\
This paper is organized as follows. In the next section we consider varying $\Lambda$ \cite{SadFar} as time-dependent dark energy and recall two dark energy models, the first is barotropic dark energy with time-dependent equation of state \cite{Khur2} and the second is Chaplygin gas model of dark energy. We give master equations with thermodynamical analysis of the solutions. Then, we review some important thermodynamics relations which affected by dynamical dark energy. In section IV we obtain the effect of dynamical dark energy on the correlation function and compare results with Peebles's power law.
\section{Dark energy model}
In the $\Lambda$-CDM model, the cosmological constant $\Lambda$ plays a role of dark energy. However,  it is not a dynamical model of dark energy (a consistent model with several observational data) and, hence, can't describe the early Universe. In order to improve this situation, we can consider a time-dependent cosmological constant $\Lambda (t)$. In that case a simplest case considered by the Ref. \cite{PourMNRAS}, now we can extend such ansatz and consider more general case including both power law and exponential form as follows,
\begin{equation}\label{Lambda}
\Lambda(t)=\Lambda_{0}(1+t^{-b_{1}}e^{-b_{2}t}),
\end{equation}
where $b_{1}$ and $b_{2}$ are free parameters of the model and can be fixed by observational data. In the case of $b_{2}=0$ we recover results of the Ref. \cite{PourMNRAS}. It is clear that $t\rightarrow\infty$ gives $\Lambda(t)=\Lambda_{0}$, where $\Lambda_{0}$ is the present value of the cosmological constant and we recover results of the $\Lambda$-CDM model. Therefore, $\Lambda(t)$ plays a role of time-dependent dark energy density $\rho$,
\begin{equation}\label{rho}
\rho=\rho_{0}(1+t^{-b_{1}}e^{-b_{2}t}),
\end{equation}
where $\rho_{0}$ is the present value of the dark energy density. The universe in this scenario described by the following FRW metric:
\begin{equation}\label{metric}
ds^{2}=-dt^{2}+a^{2}(dr^{2}+r^{2}d\Omega^{2}),
\end{equation}
where $a$ is the scale factor, and $d\Omega^{2}=d\theta^{2}+\sin^{2}\theta d\phi^{2}$. It yields to the following Friedmann equations,
\begin{equation}\label{F1}
\left(\frac{\dot{a}}{a}\right)^2=\frac{\rho}{3},
\end{equation}
and
\begin{equation}\label{F2}
2\frac{\ddot{a}}{a}+\left(\frac{\dot{a}}{a}\right)^2=p,
\end{equation}
where the dark energy density satisfies the following conservation equations:
\begin{equation}\label{conservation}
\dot{\rho}+3H(p+\rho)=0,
\end{equation}
while $p$ is the pressure and has the following equation of state:
\begin{equation}\label{EoS}
p=\omega\rho.
\end{equation}
Also, the Hubble expansion parameter can be written in terms of the scale factor $a$ as follows,
\begin{equation}\label{Hubble0}
H=\frac{\dot{a}}{a}.
\end{equation}
In order to have thermodynamical analysis of the models, we should note that the evolution rate of the dark energy entropy with the dark energy density $\rho$
is given by \cite{setare},
\begin{equation}\label{27}
{\dot{S}}_{de}=\frac{4\pi}{T}(1+\omega)\rho r_{A}^{2}(\dot{r}_{A}-Hr_{A}),
\end{equation}
where $r_{A}$ is the apparent horizon radius. Apparent horizon radius is associated with the gravitational entropy for a dynamical space-time and given by the following expression \cite{ch}:
\begin{equation}\label{28}
r_{A}=H^{-1}.
\end{equation}
Moreover, the temperature is given by,
\begin{equation}\label{T}
T=\frac{1}{2\pi r_{A}}.
\end{equation}
In this paper, we would like to consider two special cases of dark energy model which are discussed in detail with separated subsections.
\subsection{Barotropic fluid}
\begin{figure}
\begin{center}$
\begin{array}{cccc}
\includegraphics[width=95 mm]{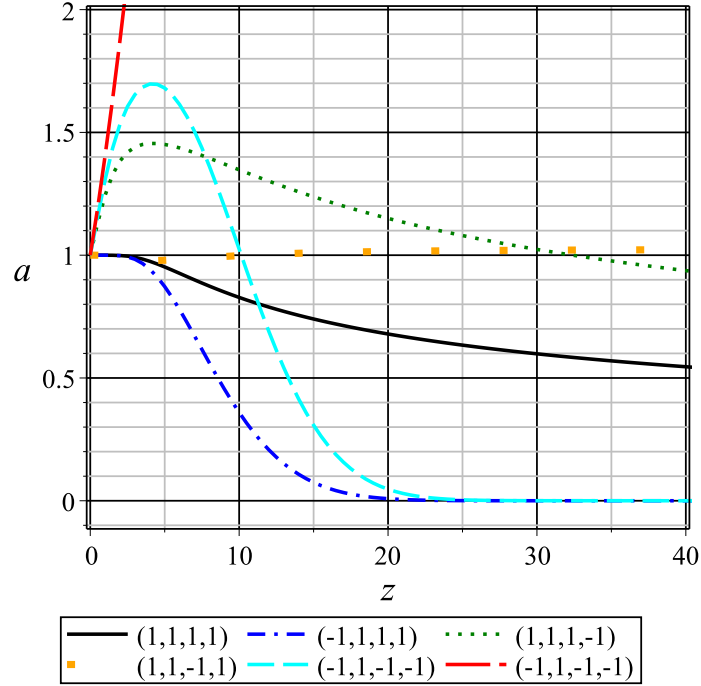}
\end{array}$
\end{center}
\caption{Typical behavior of the scale factor in terms of redshift for various values of set of parameters as ($\omega_{0},\omega_{1},b_{1},b_{2}$).}
\label{fig1}
\end{figure}

A barotropic fluid with time dependent equation of state \cite{Gor},
\begin{equation}\label{29}
\omega=\omega_{0}+\omega_{1}t,
\end{equation}
could play role of the time-dependent dark energy, where $\omega_{0}$ and $\omega_{1}$ are constants. Such time-dependent equation of state appears due to the modification of gravity \cite{noj}.\\
Now, by using the equations (\ref{rho}), (\ref{EoS}) and (\ref{29}) in the conservation equation (\ref{conservation}) one can obtain the following time-dependent Hubble parameter:
\begin{equation}\label{Hubble}
H=\frac{(b_{1}+b_{2}t)t^{-b_{1}}e^{-b_{2}t}}{3t(1+\omega_{0}+\omega_{1}t)(1+t^{-b_{1}}e^{-b_{2}t})}.
\end{equation}
Hence, one can obtain,
\begin{equation}\label{scale}
\ln{\frac{a}{a_{0}}}=\int\frac{(b_{1}+b_{2}t)t^{-b_{1}}e^{-b_{2}t}}{3t(1+\omega_{0}+\omega_{1}t)(1+t^{-b_{1}}e^{-b_{2}t})}dt.
\end{equation}
Above integral can solved numerically to obtain time dependent scale factor. We expect $a$ as increasing function of time, and it helps us to fix some free parameters of the model. We find numerical solution of the equation (\ref{scale}) and find that negative $b_{1}$ yields to decreasing scale factor which is not good result, hence we can choose positive $b_{1}$ together with
both positive or negative $b_{2}$. Also by using the results of the Ref. \cite{z} we can present our results in terms of the redshift. We expect that the scale factor be decreasing function of the redshift. We draw redshift dependent scale factor in the Fig. \ref{fig1} and find that both $b_{1}$ and $b_{2}$ should be positive while $\omega_{0}$ may be positive or negative. We denote set of free parameters as $(\omega_{0},\omega_{1},b_{1},b_{2})$ and show that the best choice may $(+,+,+,+)$ or $(-,+,+,+)$. We used unit values for all parameters to find typical behavior, while other values make no important changes in the behavior of scale factor.\\

\begin{figure}
\begin{center}$
\begin{array}{cccc}
\includegraphics[width=95 mm]{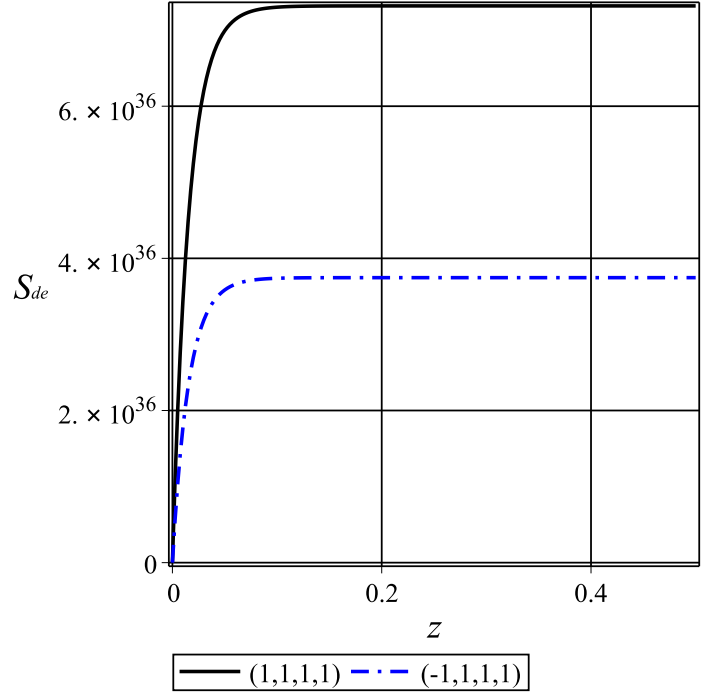}
\end{array}$
\end{center}
\caption{Typical behavior of the entropy in terms of $t$ for barotropic fluid.}
\label{fig2}
\end{figure}

Now, by exploiting relations (\ref{Hubble}) and (\ref{27}), we obtain the following differential equation:
\begin{equation}\label{27-1}
{\dot{S}}_{de}=\frac{648\rho_{0}\pi^{2}t^{3}}{(b_{1}+b_{2}t)^{5}}\left[\frac{(1+\omega_{0}+\omega_{1}t)(1+t^{-b_{1}}e^{-b_{2}t})}{b_{1}t^{-b_{1}}e^{-b_{2}t}}\right]^{4}Y(t),
\end{equation}
where we have defined,
\begin{equation}\label{Y}
Y(t)\equiv Y_{1}t^{-b_{1}}e^{-b_{2}t}+Y_{2}.
\end{equation}
with
\begin{eqnarray}\label{Y1}
Y_{1}&=&b_{2}(\omega_{1}-\frac{b_{2}}{3})t^{2}+2b_{1}(\omega_{1}-\frac{b_{2}}{3})t\nonumber\\
&-&\frac{b_{1}}{3}(b_{1}-3(1+\omega_{0})),
\end{eqnarray}
and
\begin{eqnarray}\label{Y2}
Y_{2}&=&\omega_{1}b_{2}^{2}t^{3}+(2b_{1}b_{2}\omega_{1}+(1+\omega_{0})b_{2}+\omega_{1})t^{2}\nonumber\\
&+&b_{1}(b_{1}\omega_{1}+2b_{2}(1+\omega_{0})+2\omega_{1})t\nonumber\\
&+&b_{1}(1+\omega_{0})(1+b_{1}).
\end{eqnarray}
Numerically, it is easy write the equation (\ref{27-1}) in terms of redshift \cite{z}  and solve it to confirm that entropy is increasing function of redshift which may be in agreement with the results of the Ref. \cite{zz}. In the Fig. \ref{fig2}, we can see the typical behavior of the dark energy entropy in terms of redshift.

\subsection{Chaplygin gas}
\begin{figure}
\begin{center}$
\begin{array}{cccc}
\includegraphics[width=65 mm]{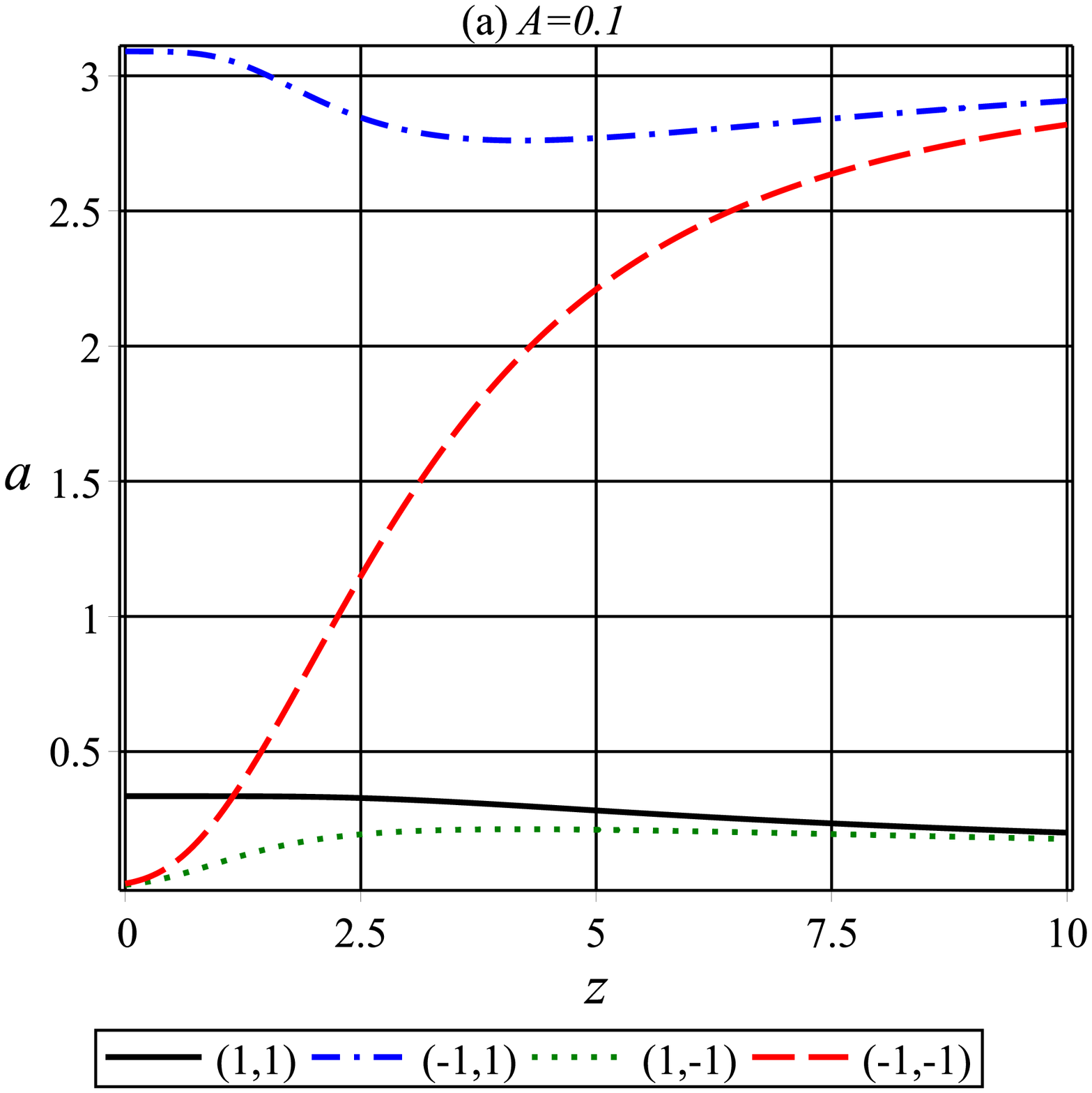}
\includegraphics[width=65 mm]{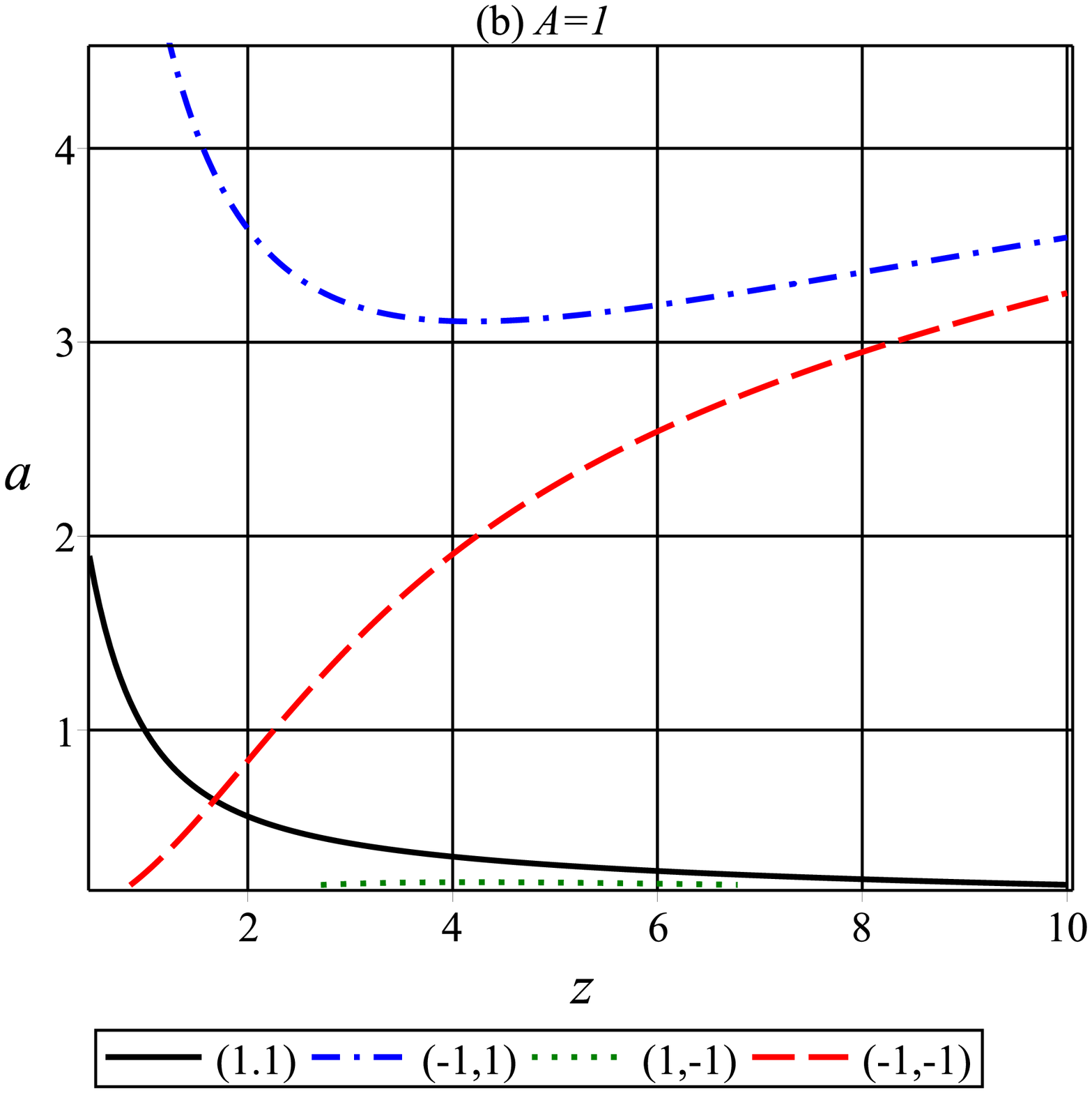}
\end{array}$
\end{center}
\caption{Typical behavior of the scale factor of Chaplygin gas model in terms of the redshift for various values of set of parameters as $(b_{1},b_{2})$.}
\label{fig3}
\end{figure}

Chaplygin gas equation of state  is given by,
\begin{equation}\label{CEoS}
\omega=-\frac{A}{\rho^{2}},
\end{equation}
where $A$ is a positive constant.
Now, by using the equations (\ref{rho}), (\ref{EoS}) and (\ref{CEoS}) in the conservation equation (\ref{conservation}), one can obtain the following time-dependent Hubble parameter:
\begin{equation}\label{Hubble2}
H=\frac{(b_{1}+b_{2}t)t^{-b_{1}}e^{-b_{2}t}}{3t\left(1-\frac{A}{(1+t^{-b_{1}}e^{-b_{2}t})^{2}}\right)(1+t^{-b_{1}}e^{-b_{2}t})}.
\end{equation}
In turn, we obtain the following scale factor,
\begin{equation}\label{scale2}
a=a_{0}\frac{e^{\frac{b_{2}t}{3}}}{\left(t^{-2b_{1}}+2t^{-b_{1}}e^{b_{2}t}+e^{2b_{2}t}-Ae^{2b^{2}t}\right)^{\frac{1}{6}}}.
\end{equation}
The above relation can easily written in terms of the redshift. Behavior of the scale factor  is depends on the values of $b_{1}$ and $b_{2}$. As illustrated by plots of the Fig. \ref{fig3}, we conclude that the positive $b_{1}$ and $b_{2}$ are the best range for the constants to have scale factor as decreasing function of the redshift. We also find that value of constant $A$ is also important and should be fixed by observational data.\\
In the Fig. \ref{fig3} (a), we can see behavior of the scale factor for $A=0.1$ and different set of parameters ($b_{1}$,$b_{2}$) as $(1,1)$, $(1,-1)$ and $(-1,1)$. In the case of positive $b_{1}$ and $b_{2}$ (solid black line of the Fig. \ref{fig3} (a)), we can see that the scale factor takes a constant value at the late time (low redshift) which means that Universe stop expanding, hence it is not good result according to the recent observational data. Other range of $b_{1}$ and $b_{2}$ yields to unexpected behavior of the scale factor.\\
The Fig. \ref{fig3} (b) describes the behavior of the scale factor for  $A=1$. In the case of positive $b_{1}$ and $b_{2}$, we can see that the scale factor is decreasing function of the redshift (increasing function of time). At the early time, it grows suddenly which shows an early inflation, then Universe expands to the current stage. As previous case, the values of  $b_{1}$ and $b_{2}$ should be only positive to give reasonable behavior.\\
Comparing with the previous case of barotropic fluid, we can see that Chaplygin gas has reasonable behavior and has better fit  with the current observational data. We also observe that both $b_{1}$ and $b_{2}$ should be positive, hence we shall use this condition in the rest of paper.\\
In the next section, we shall discuss about the entropy by using the equation (\ref{27}) for the generalized Chaplygin gas mosel.\\

\subsection{Generalized Chaplygin gas}
\begin{figure}
\begin{center}$
\begin{array}{cccc}
\includegraphics[width=75 mm]{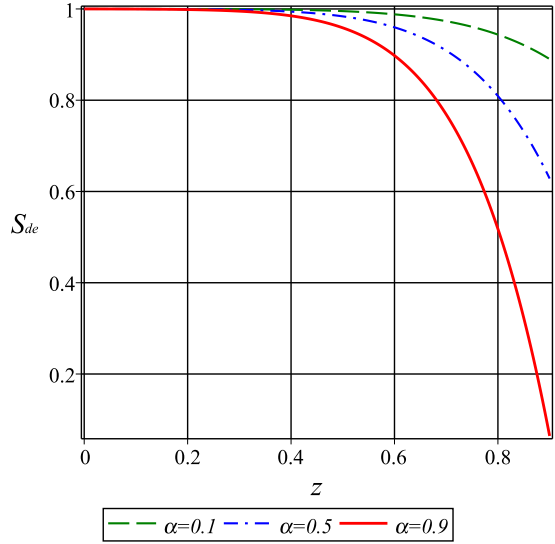}
\end{array}$
\end{center}
\caption{Typical behavior of the entropy in terms of $z$ for GCG with $A=b_{1}=b_{2}=1$.}
\label{fig4}
\end{figure}

The generalized Chaplygin gas equation of state is given by,
\begin{equation}\label{GCEoS}
\omega=-\frac{A}{\rho^{1+\alpha}},
\end{equation}
where $0<\alpha\leq1$ is a constant. By setting the particular case of $\alpha=1$, the Chaplygin gas considered in the previous subsection recovered.\\
Now, by using the equations (\ref{rho}), (\ref{EoS}) and (\ref{GCEoS}) in the conservation equation (\ref{conservation}) we obtain the following time-dependent Hubble parameter,
\begin{equation}\label{Hubble3}
H={\frac {(3t)^{-1} \left( {t}^{b_{1}}{{\rm e}^{b_{2}t}}+1 \right) ^{1+\alpha}
 \left( b_{2}t+b_{1} \right) }{ \left( {t}^{b_{1}}{{\rm e}^{b_{2}t
}}+1 \right) ^{1+\alpha}-\frac{A{t}^{b_{1} \left( 2+\alpha \right) }}{{{\rm e}^{-b_{2}t
 \left( 2+\alpha \right) }}}+{{\rm e}^{b_{2}t}} \left( {t}^{b_{1}}{{\rm e}^{b_{2}t}
}+1 \right) ^{1+\alpha}{t}^{b_{1}}  }}.
\end{equation}
In this case,  the scale factor (\ref{scale2}) generalized to the following expression:
\begin{equation}\label{scale3}
a=a_{0}\frac{e^{\frac{b_{2}t}{3}}
\left((1+t^{-b_{1}}e^{-b_{2}t})^{\alpha}-\frac{A}{1+t^{-b_{1}}e^{-b_{2}t}}\right)^{-\frac{1}{3(1+\alpha)}}}
{(1+t^{-b_{1}}e^{-b_{2}t})^{-\frac{\alpha}{3(1+\alpha)}}\left(t^{-b_{1}}+e^{b_{2}t}\right)^{\frac{1}{3}}}.
\end{equation}
Also, in this case,  the   differential equation for the entropy is given by,
\begin{eqnarray}\label{27-1-1}
\dot{S_{de}}&=&-\frac{216\pi^2\big(A-\rho_0^{\alpha+1}(1+t^{-b_1}e^{-b_2t})^{\alpha+1}\big)^4}
{\rho_0^{5\alpha+6}(b_1+b_2t)^5(1+t^{-b_1}e^{-b_2t})^{5\alpha+1}}\nonumber\\
&\times& t^{5b_1+5}e^{5b_2t}\bigl(Z_1(t)+Z_2(t)\bigr),
\end{eqnarray}
where we have defined
\begin{eqnarray}
Z_1(t)&=&\biggl(4\rho_0^{\alpha+1}(1+t^{-b_1}e^{-b_2t})^{\alpha+1}+3A\alpha\biggr)\nonumber\\
&\times&\rho_0^{2}t^{-2b_1-2}e^{-2b_2t}(b_1+b_2t)^2,
\end{eqnarray}
and
\begin{eqnarray}
Z_2(t)&=&\biggl(3A\rho_0(1+t^{-b_1}e^{-b_2t})-3\rho_0^{\alpha+2}(1+t^{-b_1}e^{-b_2t})^{\alpha+2}\biggr)\nonumber\\
&\times&\rho_0t^{-b_1-2}e^{-b_2t}\bigl((b_1+b_2t)^2+b_1\bigr).
\end{eqnarray}
Numerical  analysis of the   equation (\ref{27-1-1}) shows that the dark energy entropy is
an increasing function of time (decreasing function of the redshift), hence the second law of thermodynamics holds in this model. In the Fig. \ref{fig4}, we can see the typical behavior of the dark energy entropy in terms of the redshift.

\section{Thermodynamics of galaxy clustering}
The partition function for $N$ particle (galaxy) system is given by (\cite{ahm02}):
\begin{equation}\label{1}
Z(T,V)=\frac{1}{\lambda^{3N}N!}\int d^{3N}pd^{3N}r X,
\end{equation}
where $X$ is defined as
\begin{equation}\label{1-1}
X\equiv \exp\biggl(-\frac{1}{T}\biggl[\sum_{i=1}^{N}\frac{p_{i}^2}{2m}+\Phi(r_{1}, r_{2}, r_{3}, \dots, r_{N})\biggr]\biggr),
\end{equation}
Here, units are chosen such that the Boltzmann constant becomes one. Upon performing integration over momentum space, the partition function (\ref{1}) leads to the following expression:
\begin{equation}\label{zn}
Z_N(T,V)=\frac{1}{N!}\left(\frac{2\pi mT}{\lambda^2}\right)^{3N/2}Q_N(T,V),
\end{equation}
where configurational integral is expressed by
\begin{equation}\label{q1}
Q_{N}(T,V)=\int....\int \prod_{1\le i<j\le N} (f_{ij}+1)d^{3N}r.
\end{equation}
Here, two-particle function $f_{ij}$ can be expressed in terms of potential energy as follows,
\begin{equation}\label{fun}
f_{ij}=e^{-\frac{\Phi(r_{ij})}{T}}-1,
\end{equation}
which obviously vanishes in the absence of interactions \cite{ahm02}.
In order to remove divergence, one can introduce the softening parameter $\epsilon$ to the Newton's potential energy  \cite{ahm02},
\begin{equation}\label{5}
\Phi(r_{ij})=-\frac{Gm^2}{(r_{ij}^2+\epsilon^2)^{1/2}}.
\end{equation}
Also, the modified potential due to the dark energy (dynamical cosmological constant) is given by \cite{pe},
\begin{equation}\label{6}
\Phi(r_{ij})=-\frac{Gm^2}{(r_{ij}^2+\epsilon^2)^{1/2}}-\frac{\Lambda(t) r_{ij}^2}{6}.
\end{equation}
The case of constant $\Lambda$, in agreement with $\Lambda$-CDM model, has been studied already by \cite{main}. Moreover, time-dependent cosmological constant $\Lambda(t)$ with inverse square potential has been recently studied by \cite{Capozziello3}. In this case,
the two-particle function takes the following form:
\begin{eqnarray}\label{7}
f_{ij}=\frac{Gm^{2}}{(r_{ij}^{2}+\epsilon^{2})^{1/2}T}+\frac{\Lambda(t) r_{ij}^2}{6T},
\end{eqnarray}
where $\Lambda(t)$ is given by the equation (\ref{Lambda}).
Now, we calculate the configurational integral for $N$ particle case as
\begin{equation}\label{qn}
Q_{N}(T,V)=V^N\left[1+\beta x\right]^{N-1},
\end{equation}
where
\begin{equation}\label{qn-1}
\beta\equiv\sqrt{1+\frac{\epsilon^2}{R_1^2}} +\frac{\epsilon^2}{R_1^2} \log \frac{{\epsilon/R_1}}{\left[ 1+\sqrt{1+\frac{\epsilon^2}{R_1^2}}\right]}+\frac{\Lambda(t) R_1^3}{15 Gm^2},
\end{equation}
with
\begin{equation}\label{qn-1-1}
R_{1}\sim \varrho^{-1/3}\sim (\bar N/V)^{-1/3}
\end{equation}
and
\begin{equation}\label{sc}
x\equiv\frac{3}{2}\left(\frac{ Gm^{2}}{T}\right)^3\varrho.
\end{equation}
Utilizing above expression of configurational integral, the gravitational partition function has following expression:
\begin{eqnarray}\label{11}
&&Z_N(T,V)=\frac{V^{N}}{N!}\left(\frac{2\pi mT}{\lambda^2}\right)^{3N/2}\nonumber\\
&&\times\left[1+ \beta x\right]^{N-1}.
\end{eqnarray}
Now, with the help of partition function, we are able to  calculate thermodynamic quantities like Helmholtz free energy which is given by
\begin{eqnarray}
F&=&-T\ln Z_{N}(T,V),\nonumber\\
&=&NT\log\left(\frac{N}{V}T^{-3/2}\right)-NT -NT\log\big[1+\beta x\big]\nonumber\\
& -& \frac{3}{2}NT\log\left(\frac{2\pi m}{\lambda^2}\right).
\end{eqnarray}
Then, we can obtain the entropy via
\begin{eqnarray}\label{1919}
S&=& -\left(\frac{\partial F}{\partial T}\right)_{N,V},\nonumber\\
&=& N\log\left(\frac{V}{N}T^{3/2}\right)+N\log [1+\beta x]-3N\frac{\beta x}{1+\beta x}\nonumber\\
&+&\frac{5}{2}N+\frac{3}{2}N\log\left(\frac{2\pi m}{\lambda^2}\right).
\end{eqnarray}
One can obtain internal energy of a system of galaxies as
\begin{equation}\label{20}
U=F+TS=\frac{3}{2}NT\left[1-2 \frac{\beta x}{1+\beta x}\right].
\end{equation}
The specific heat can be calculated via the following relation:
\begin{equation}\label{21}
C=T\left(\frac{\partial S}{\partial T}\right)_{V}.
\end{equation}
Grand canonical partition function in terms of canonical partition function is given by,
\begin{eqnarray}\label{g}
Z_{G}(T,V,z)=\sum_{N=0}^{\infty}z^{N}Z_{N}(V,T),
\end{eqnarray}
where
\begin{eqnarray}\label{23}
z=e^{\frac{\mu}{T}}
\end{eqnarray}
is called  activity and
\begin{eqnarray}\label{24}
\mu &=& \biggl(\frac{\partial F}{\partial N}\biggr)_{V,T},\nonumber\\
&=&{T}\log \left(\frac{N}{V} T^{-3/2}\right)+{T}\log\left[1- \frac{\beta x}{1+\beta x}\right]\nonumber\\
&-& \frac{3}{2}{T}\log\left(\frac{2\pi m}{\lambda^2}\right)-\frac{\beta x}{1+\beta x}{T}
\end{eqnarray}
is chemical potential.
By using the relations (\ref{g}), (\ref{11}) and  (\ref{24}) one can obtain the probability of finding $N$ particles in volume $V$ as follow,
\begin{eqnarray}\label{26}
P&=&\frac{e^{\frac{N\mu}{T}}Z_{N}(V,T)}{Z_{G}(T,V,z)},\nonumber\\
&=& \frac{\bar{N}}{N!}\left(   \bar N  + N \beta x  \right)^{N-1}\left( 1+\beta x\right)^{-N}
e^{{\frac{- \bar N  - N \beta x }{1+\beta x}}}.
\end{eqnarray}
Now, we can study the total entropy due to
the dark energy ($S_{de}$) and clustering of galaxies (given by the equation (\ref{1919})),
\begin{eqnarray}
S_{tot}=S+S_{de},
\end{eqnarray}
which is different for the barotropic or Chaplygin gas dark energy. The first part yields to a constant after long time, hence the second part ($S_{de}$) is dominant which is illustrated by Figs. \ref{fig2} and \ref{fig4}. Therefore, the only model where the second law of thermodynamics is valid given by the GCG.

\section{Effect of the dynamical dark energy on the correlation function}
\begin{figure}
\begin{center}$
\begin{array}{cccc}
\includegraphics[width=75 mm]{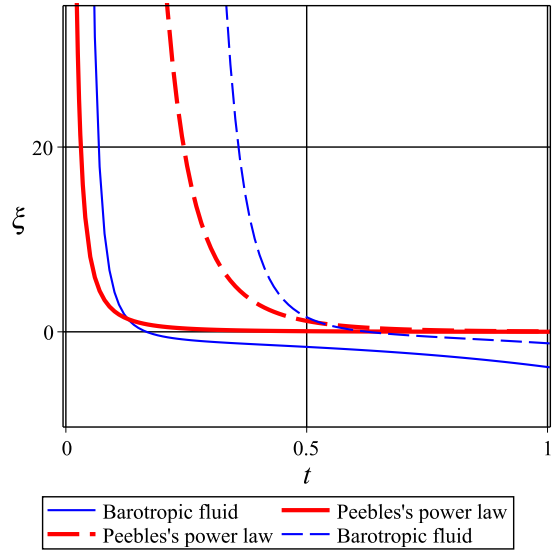}
\end{array}$
\end{center}
\caption{Typical behavior of the correlation function in terms of $t$. Solid lines drawn for $\omega_{0}=\omega_{1}=1$ and $b_{1}=b_{2}=1$, while dashed lines drawn for $\omega_{0}=-1$, $\omega_{1}=1$, and $b_{1}=b_{2}=1$. All other parameters sets as unity.}
\label{fig7}
\end{figure}

In order to find the suitable model in agreement with simulations, we need to calculate the
correlation function.
As we know, the two-point correlation function $\xi(r)$ of the clustering of galaxies obeys power law $\xi(r)=r^{-1.8}$ \cite{pee80} which has been confirmed from $N$-body simulation \cite{sut90}. Now, we would like to study the effect of dynamical dark energy as barotropic and Chaplygin gas models on the correlation function.\\
The main equation of internal energy is given by \cite{sas00,ham16},
\begin{equation}\label{42}
U=\frac{3}{2}NT-\frac{N^{2}}{2V}\int_{V}\Phi(r,T)\xi(r)4\pi r^2dr.
\end{equation}
Here, Boltzmann's constant is unit.
Then, by using the interaction potential $\Phi$ and the clustering parameter,
\begin{equation}\label{44}
\mathcal{B}=\frac{Gm^2 N}{6TV}\int\left[\frac{\xi(r)}{\sqrt{\epsilon^2+r^2}}+\frac{\rho r^3}{6}\right]dV,
\end{equation}
we can obtain correlation function $\xi(r)$. In that case one can obtain,
\begin{equation}\label{49}
\xi(r)=\frac{9{\mathcal{B}}^{2}TV}{2\pi Gm^2N}\frac{1}{r^2}\bigl(1+\frac{\epsilon^2}{2r^2}-\frac{\rho r^3}{6}\bigr),
\end{equation}
where $\rho$ is given by the equation (\ref{rho}). Hence, it is different for Barotropic fluid and Chaplygin gas.\\
First of all we consider barotropic fluid equation of state and study the equation (\ref{49}).
In the Fig. \ref{fig7} we give typical behavior of correlation function (\ref{49}) for barotropic fluid (see blue lines) and compare it with Peebles's power law (see red lines). We can see general agreement for correlation function, specially for the small $t$. It means the Barotropic fluid could describe the early Universe.\\

\begin{figure}
\begin{center}$
\begin{array}{cccc}
\includegraphics[width=75 mm]{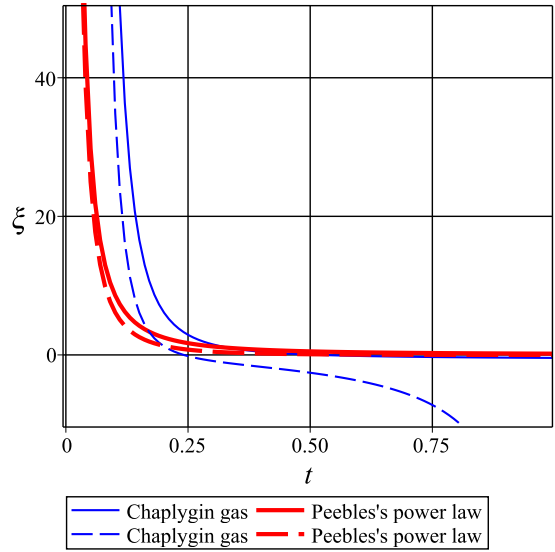}
\end{array}$
\end{center}
\caption{Typical behavior of the the correlation function in terms of $t$ with $A=1$. Solid lines drawn for $b_{1}=b_{2}=1$, while dashed lines drawn for $b_{1}=1$, and $b_{2}=-1$.}
\label{fig8}
\end{figure}

Then, we consider Chaplygin gas equation of state and give numerical study of the equation (\ref{49}).
In the Fig. \ref{fig8} we give typical behavior of correlation function (\ref{49}) for the Chaplygin gas equation of state (see blue lines) and compare it with Peebles's power law (see red lines). We can see general agreement for correlation function more than barotropic fluid (specially for positive  $b_{1}$ and $b_{2}$). We can see the good agreement behavior for the large $t$. We can conclude that the Chaplygin gas equation of state may describe the late time Universe.

\section{Conclusion}
In this paper, we have assumed  that varying cosmological constant corresponds to a time-dependent dark energy. There are various dynamical dark energy models in order to explain fine-tuning  and the cosmic coincidence problems. Two such models of our interest are the  barotropic fluid  and the  Chaplygin gas model. In order to do comparative analysis, we have studied galaxy clustering in presence of these dark energy models.   In both cases, we have found a scale factor for various values of the model parameters. In the case of generalized Chaplygin gas and  barotropic fluid, we studied the behavior of the dark energy entropy with time and found that it is an increasing function in both the cases. Numerically, we have shown that the value of dark energy entropy in the case of generalized Chaplygin gas is greater than the case of barotropic fluid. Moreover, we have investigated thermodynamical entities like Helmholtz free energy, entropy, internal energy,  specific heat and chemical potential. We have calculated probability and partition function as well. We could see the effect of each models on the clustering parameter. Hence, we can write total entropy of the system including the dark energy and clustering of galaxies. We have shown that dark energy entropy is dominant and clustering of galaxies entropy are negligible. Finally, we calculated the effect of barotropic and Chaplygin gas models on the correlation function and compare results with Peebles's power law. We found that the Chaplygin gas gives better fit to the original Peebles's power law than the barotropic fluid. However we can conclude that the barotropic fluid may describe the early Universe while Chaplygin gas describe the late time Universe. Therefore, as the Chaplygin gas equation of state can be obtained from string theory, it may be an observational validation for the string theory. We should note that here we only consider a toy model including  generalized Chaplygin gas while there are other better models like modified or extended  Chaplygin gas.
It would be interesting to consider other models of Chaplygin gas like extended Chaplygin gas \cite{ECG4, ECG5, ECG6} to investigate the clustering and cosmic energy equation of galaxies thermodynamics. These are subject of further investigations.

\end{document}